\documentclass[journal=jacsat,manuscript=article]{achemso}
\usepackage{siunitx}
\usepackage{graphicx} 
\usepackage[utf8]{inputenc}
\usepackage[T1]{fontenc}
\usepackage{amsmath}
\usepackage{amsfonts}
\usepackage{amssymb}
\usepackage{array}
\usepackage[version=3]{mhchem} 
\usepackage{braket}
\usepackage{bm}
\usepackage{xr}
\usepackage{xspace}
\usepackage{multirow}

\usepackage{cleveref}
\usepackage{titlesec}

\usepackage{xspace} 
\usepackage{caption}
\usepackage{subcaption}
\usepackage{tikz}
\usetikzlibrary{arrows, arrows.meta}
\usepackage{comment}
\usepackage{newtxtext}
\usepackage{mathtools}
\setkeys{acs}{articletitle = true}

\DeclareUnicodeCharacter{0301}{\'{e}}

\usepackage{xr-hyper}
\makeatletter
\newcommand*{\addFileDependency}[1]{
 \typeout{(#1)}
 \@addtofilelist{#1}
 \IfFileExists{#1}{}{\typeout{No file #1.}}
}
\makeatother

\newcommand*{\myexternaldocument}[1]{
 \externaldocument{#1}
 \addFileDependency{#1.tex}
 \addFileDependency{#1.aux}
}

\myexternaldocument{SI}

\newcommand*{\sm}{SI\xspace}

\author{Chiara Sepali}
\affiliation{Scuola Normale Superiore, 56126 Pisa, Italy}
\author{Linda Goletto}
\affiliation{Scuola Normale Superiore, 56126 Pisa, Italy}
\author{Piero Lafiosca}
\affiliation{Scuola Normale Superiore, 56126 Pisa, Italy}
\author{Matteo Rinaldi}
\affiliation{Scuola Normale Superiore, 56126 Pisa, Italy}
\author{Tommaso Giovannini}
\affiliation{Department of Physics, University of Rome Tor Vergata, Via della Ricerca Scientifica 1, 00133, Rome, Italy}
\author{Chiara Cappelli}
\affiliation{Scuola Normale Superiore, 56126 Pisa, Italy}
\email{chiara.cappelli@sns.it}

\title{Fully Polarizable Multiconfigurational Self-Consistent Field/Fluctuating Charge Approach}

\abbreviations{IR,NMR,UV}

\begin{document}




\begin{abstract}
A multiscale model based on the coupling of the multiconfigurational self-consistent field (MCSCF) method and the classical atomistic polarizable Fluctuating Charges (FQ) force field is presented. 
The resulting MCSCF/FQ approach is validated by exploiting the CASSCF scheme through application to compute vertical excitation energies of formaldehyde and para-nitroaniline in aqueous solution. The procedure is integrated with molecular dynamics simulations to capture solute's conformational changes and the dynamic aspects of solvation. Comparative analysis with alternative solvent models, gas-phase calculations, and experimental data provides insights into the model's accuracy in reproducing solute-solvent molecular interactions and spectral signals. 
\end{abstract}


\newpage

\section*{Introduction}

Multiconfiguration self-consistent field (MCSCF) techniques offer a useful and widely used alternative to single reference quantum mechanical (QM) methods, that often fail when applied to highly correlated chemical systems \cite{schmidt1998construction,szalay2012multiconfiguration} such as carotenoids \cite{spezia2004dft} transition metal complexes, \cite{neese2007advanced} and especially photochemical processes.\cite{curchod2018ab,crespo2018recent} 

In MCSCF, wavefunctions are expressed as the superposition of multiple electronic configurations; molecular orbitals (MOs) and configuration interaction (CI) coefficients of electronic configurations are variationally optimized, and MOs are categorized into inactive, active, and virtual spaces. 
\cite{grein1971multiconfiguration,roos1980complete,malmqvist1990restricted,werner1985second,olsen2011casscf} 
Molecular states can be optimized using either a state-specific (SS) approach or a state-average (SA) scheme, in which a weighted average of the energy of all selected states is minimized over a set of common orbitals.\cite{ruedenberg1979mcscf,werner1981quadratically,docken1972lih}
Currently, the most widespread approach for performing MCSCF calculations consists of resorting to the complete active space self-consistent field (CASSCF) approach, where the wavefunction expansion includes all potential configuration state functions within a given set of active orbitals.\cite{roos1980complete,olsen2011casscf} The scalability of this method has advanced considerably, with the largest calculations performed using an active space of 22 electrons in 22 active orbitals. \cite{vogiatzis2017pushing}
Also, its accuracy is excellent, especially after incorporating dynamic correlation effects, for instance by using second-order perturbation theory based on a CASSCF reference state, i.e. the CASPT2 method. All this makes the approach a gold standard for accurate ab initio calculations of excited-state energies, properties,\cite{andersson1990second,andersson1992second,schreiber2008benchmarks,silva2010benchmarks,lischka2018multireference,gonzalez2012progress,barneschi_2023_assessment} and especially non-adiabatic excited state dynamics.\cite{lischka2018multireference,gonzalez2012progress,mai2018nonadiabatic} 


The MCSCF methods, when applied to molecular systems in the gas phase, are highly developed and thoroughly validated. However, most excited-state photophysical and photochemical processes occur in the condensed phase, which is why methods have been developed to couple an accurate MCSCF description with (atomistic or continuum) classical or quantum embedding methods, in a multiscale fashion.\cite{giovannini2023continuum,mennucci2019multiscale,tomasi2005quantum,olsen2011molecular,warshel1976theoretical,wesolowski1993frozen,marrazzini2021multilevel} 

Multiscale hybrid QM/Classical models constitute an efficient way to study systems in the condensed phase. \cite{warshel1976theoretical,senn2009qm,lin2007qm,tomasi2005quantum,morzan2018spectroscopy} 
In these approaches, the target is accurately treated at the QM level, while the environment is described with classical physics. This allows for taking into account the effects of the environment on the target's properties in a computationally efficient way. 
%
%
%
%
%
The MCSCF method has been integrated with the Polarizable Continuum Model (PCM) in an SS framework,\cite{cammi2002second,cossi1999direct,aguilar1993nonequilibrium,amovilli1998mcscf,cammi2003multiconfigurational} and extended to an SA formulation to study conical intersections and photoreaction energy pathways of solvated systems.\cite{song2022state} A bunch of polarizable atomistic models such as Polarizable Embedding (PE), MMPol, and AMOEBA \cite{hedegaard2014polarizable,hedegaard2013multi,list2013unified,song2023state,li2015polarizable,song2024polarizable} have been coupled with MCSCF wave functions using a variety of strategies, including SS or SA approaches, or hybrid methods in which the solvent polarization is converged to the solute SA density and an SS correction is implemented to refine energy values.\cite{li2015polarizable}



In this work, we propose for the first time the coupling of an MCSCF wavefunction (within a CASSCF formalism) and the polarizable Fluctuating Charge (FQ) force field,\cite{rick1994dynamical} in a QM/MM fashion.\cite{cappelli2016integrated,giovannini2020molecular,giovannini2020theory,gomez2023multiple}
FQ is a polarizable classical force field, where polarization effects are modeled by assigning to each atom belonging to the classical portion (i.e. to the solvent) a charge ($q$) that is not fixed, like in standard classical force field, but can vary according to the electronegativity equalization principle (EEP).\cite{sanderson1951interpretation} In this way, a charge flow occurs when two atoms have different chemical potentials.\cite{rick1994dynamical} 
%
When coupled to DFT or TD-DFT Hamiltonians, FQ has been demonstrated to yield excellent results in the simulation of various spectral properties of aqueous and non-aqueous solutions \cite{giovannini2020molecular,giovannini2020theory,gomez2023multiple,ambrosetti2021quantum,giovannini2019quantum} 


The paper is organized as follows. First, the theoretical foundations of MCSCF and FQ are recalled, and then the MCSCF/FQ coupling is discussed.
After a detailed description of the computational protocol, the method is validated by simulating vertical excitation energies of formaldehyde and para-nitroaniline in aqueous solution. Formaldehyde has been selected for its compact size and is used to set the computational protocol, by examining variations in the basis set, starting orbitals, active space selection, and solvent modeling (from continuum to atomistic methods).
Para-nitroaniline is a prototypical push-pull system with a $\pi$-conjugated link between electron-donor and electron-acceptor groups. Its optical properties, which are significantly influenced by solvation, make it an ideal system for showing the performance of the newly developed method. \cite{kosenkov2011solvent,cammi2003multiconfigurational,makwani2009solvent,giovannini2019electronic} Vertical excitation energies of PNA in aqueous solution are discussed by highlighting the role of dynamical solvent fluctuations and finally compared with experimental data. Conclusion and future perspectives end the paper. 

\section{Theory}

This section presents the coupling of MCSCF with the FQ polarizable force field. Initially, the main features of MCSCF (\cref{sec:cas}) and FQ (\cref{sec:fq}) are summarized to provide a background for the description of their coupling, which is presented in \cref{sec:coupling}. The integration of FQ with MCSCF is challenging, due to the intrinsic complexity of multireference approaches. In this work, MCSCF (CASSCF formulation) is integrated with FQ by exploiting a State-specific (SS) approach, where each solute's state is optimized separately and interacts with the polarizable solvent. This strategy appears reasonable for calculating excited states' energies and properties. 
However, a state-average (SA) scheme should be preferred to study phenomena involving different excited states, such as conical intersections, photochemistry, and excited state dynamics.\cite{song2023state} The extension of MCSCF/FQ to an SA formalism is not a trivial extension of the SS formulation, therefore it will be treated in future works.

\subsection{The MCSCF approach}\label{sec:cas}

The MCSCF wavefunction is expressed as a linear combination of Slater determinants $\ket{\phi_m}$ built over a common set of molecular orbitals (MOs), which in turn are expressed as a linear combination of atomic orbitals (AOs). \cite{roos1992multiconfigurational} By following the notation of Roos,\cite{roos1992multiconfigurational} the state of interest $\ket{\Psi_0}$ and the excited states $\ket{\Psi_K}$ assume the following form:
\begin{equation}
    \ket{\Psi_0} = \sum_m C_{m0} \ket{\phi_m}, \quad \ket{\Psi_K} = \sum_m C_{mK} \ket{\phi_m}
\end{equation}
where $C_{m0}$ and $C_{mK}$ indicate CI coefficients. In MCSCF, $\ket{\Psi_0}$ is variationally optimized with respect to $C_{m0}$ and MOs coefficients. To this end, the MCSCF wavefunction $\ket{\Psi_0}$ is conveniently expressed as follows:\cite{roos1992multiconfigurational}
\begin{equation}\label{eq:trans_wf}
    \ket{\Psi_0} = \text{exp} (\hat{T})\text{exp} (\hat{S}) \ket{0}  
\end{equation}
where $\ket{0}$ is a reference state, while $\text{exp}(\hat{T})$ and $\text{exp}(\hat{S})$ are unitary transformations among MOs and states, respectively. $\hat{T}$ and $\hat{S}$ are anti-Hermitian operators which read:
\begin{equation}
    \hat{T} = \sum_{pq} T_{pq} \hat{E}_{pq}; \quad \hat{S} = \sum_{K \ne 0} S_{K0} (\ket{\Psi_K} \bra{\Psi_0} - \ket{\Psi_0} \bra{\Psi_K}) 
\end{equation}
where $\hat{E}_{pq}$ is the singlet excitation operator for generic $p,q$ MOs , while $T_{pq}$ and $S_{K0}$ are variational parameters. 

The energy of $\ket{\Psi_0}$ can thus be written as:
\begin{equation}\label{eq:energy_cas_exp}
    E_\mathrm{MCSCF} = \bra{0} \text{exp}(-\hat{S}) \text{exp}(-\hat{T}) \hat{H} \text{exp}(\hat{T}) \text{exp}(\hat{S}) \ket{0} 
\end{equation}
where $\hat{H}$ indicates the Hamiltonian, which reads (in second quantization):
\begin{equation}\label{eq:hamiltonian-sq}
    \hat{H} = \sum_{pq} h_{pq} \hat{E}_{pq} + \frac{1}{2} \sum_{pqrs} g_{pqrs} (\hat{E}_{pq} \hat{E}_{rs} - \delta_{rq} \hat{E}_{ps}) 
\end{equation}
$h_{pq}$ and $g_{pqrs}$ are one- and two-electron integrals in the MO basis, respectively. \Cref{eq:energy_cas} can be re-written as a function of $\mathbf{D}$ and $\mathbf{P}$, i.e. the one-particle and two-particle density matrices associated with $\ket{\Psi_0}$:\cite{helgaker2013molecular}
\begin{equation}\label{eq:energy_cas}
E_\mathrm{MCSCF}(\mathbf{D}, \mathbf{P}) = \sum_{pq}D_{pq} h_{pq} + \sum_{pqrs} P_{pqrs} g_{pqrs}
\end{equation}
By applying the Baker-Campbell-Hausdorff (BCH) expansion, the electronic MCSCF gradients and Hessians are obtained by differentiating the energy with respect to the wavefunction parameters. The stationarity of the energy functional is guaranteed when gradients with respect to CI coefficients ($g_K^{(c)}$) and MOs coefficients ($g_{rs}^{(o)}$) satisfy the two following conditions:
\begin{align}
    g_{K}^{(c)} & = 2\braket{\Psi_0|\hat{H}|\Psi_K} = 0 \label{eq:ci_grad} \\
    g_{rs}^{(o)} &= \braket{\Psi_0| [\hat{H}, \hat{E}_{rs}^-]|\Psi_0} = 2 \braket{\Psi_0| \hat{H} \hat{E}_{rs}^-|\Psi_0} = 2 \braket{\Psi_0|\hat{H}|rs} = 0\label{eq:orb_grad}
\end{align}
where $\hat{E}_{rs}^- = \hat{E}_{rs} -\hat{E}_{sr}$ and $\ket{rs} = \hat{E}_{rs}^-\ket{\Psi_0}$ are the so-called Brillouin states.\cite{kreplin2020mcscf} \Cref{eq:ci_grad} is equivalent to the secular problem $(\mathbf{H}-E\mathbf{1})\mathbf{C} = \mathbf{0}$ and \cref{eq:orb_grad} is the Brillouin-Levy-Berthier (BLB) theorem, also known as the \textit{Extended Brillouin Theorem}. \cite{roos1992multiconfigurational,levy1968generalized,levy1969generalized} 

The optimization of MCSCF wavefunctions is generally carried out using first-order  or second-order methods (see Refs. \citenum{schmidt1998construction,kreplin2019second} and references therein)
depending on whether the MOs and CI coefficients are optimized alternatively or simultaneously, respectively. 
In this work, we resort to a first-order approach implemented in OpenMolcas \cite{aquilante2020modern,li2023openmolcas}, in particular by adopting the Super-CI (SCI) approach\cite{roos1980complete1,roos1980complete,malmqvist1990restricted,siegbahn1981complete} based on the BLB theorem. \cite{levy1968generalized,levy1969generalized}  
SCI can be adopted in combination with the most common method for MCSCF calculations, i.e., the complete active space self-consistent field (CASSCF) approach, where the wavefunction expansion includes all potential configuration state functions within a given set of active orbitals. \cite{roos1980complete,olsen2011casscf}
CASSCF is the method employed in the present study.

In SCI, the CI coefficients are calculated by solving the secular problem (see \cref{eq:ci_grad}) for a given set of MOs. The improved SCI wavefunction is then set up as a superposition of $\ket{\Psi_0}$ and all the Brillouin states corresponding to non-redundant orbital rotations as follows: \cite{roos1980complete1}
\begin{equation}
    \ket{\mathrm{SCI}} = \ket{\Psi_0} + \sum_{p>q} T_{pq} \hat{E}^-_{pq}  \ket{\Psi_0} = \ket{\Psi_0} + \sum_{p>q} T_{pq}\ket{pq}
\end{equation}
Therefore, the $T_{pq}$ coefficients are determined by solving the generalized eigenvalue equation: \cite{roos1980complete1}
\begin{equation}\label{eq:SCI}
\mathbf{HT} = E_\mathrm{SCI}\overline{\mathbf{S}}\mathbf{T}
\end{equation}
where $H_{rs,pq} = \braket{rs|\hat{H}|pq}$ is the Hamiltonian matrix and $\overline{\mathbf{S}}$ is the overlap matrix in the basis of $\ket{\Psi_0}$ and Brillouin states.\cite{kreplin2020mcscf} Because calculating the matrix elements between the Brillouin states $\braket{rs|\hat{H}|pq}$ is extremely expensive, alternative schemes have been proposed and amply tested in the literature.\cite{roos1980complete1,roos1980complete,siegbahn1981complete} 

\subsection{The Fluctuating Charges (FQ) force field}\label{sec:fq}

The FQ force field\cite{rick1994dynamical,rick1995fluctuating} endows each atom with a charge $q$ which dynamically adapts to the external potential. The total Lagrangian functional of the FQ model is obtained from a second-order Taylor expansion of the energy with respect to charges.\cite{rick1994dynamical} For a system composed of various molecules, it is generally written as follows:\cite{rick1994dynamical}
\begin{equation}
\begin{aligned}\label{eq:funct}
E_{\text{FQ
}} & = \sum_{i \alpha} \text{q}_{i \alpha} \chi_{i \alpha} + \frac{1}{2} \sum_{i \alpha} \sum_{j \beta} \text{q}_{i \alpha} \text{T}_{i \alpha, j \beta}^{\text{qq}} \text{q}_{j \beta} + \sum_{\alpha} \left[ \lambda_\alpha \sum_i \text{q}_{i \alpha} - \text{Q}_{\alpha} \right]
\end{aligned}
\end{equation}
where ($i, j$) and ($\alpha, \beta$) indices run over FQ atoms and molecules, respectively. $\chi_{i\alpha}$ indicates the atomic electronegativity, while 
$\text{T}_{i\alpha,j\beta}^{\text{qq}}$ is the charge-charge interaction kernel. To prevent the so-called ``polarization catastrophe'',\cite{thole1981molecular} the Ohno kernel is adopted.\cite{ohno1964some} 
Its diagonal elements $T_{i\alpha,i\alpha}^{qq}$ are expressed in terms of the atomic chemical hardness $\eta_{i\alpha}$. A set of Lagrangian multipliers $\lambda_\alpha$ are introduced in \cref{eq:funct} to constrain the total charge of each molecule to $Q_\alpha$. This is done to avoid artifacts arising from unphysical charge transfer between different molecules.\cite{chen2009dissociation} As a result of its formulation, FQ relies on two atomic parameters, $\chi_{i\alpha}$ and $\eta_{i\alpha}$, which can be rigorously defined in the framework of Conceptual Density Functional Theory.\cite{chelli2002transferable,mortier1985electronegativity} 

Charge exchange between atoms is driven by the Electronegativity Equalization Principle,\cite{sanderson1951interpretation} which is satisfied by imposing the stationarity conditions of the Lagrangian functional with respect to atomic charges and multipliers. This translates into solving the following linear system: \cite{cappelli2016integrated,giovannini2020molecular}
\begin{equation}\label{eq:fq-linear}
\left(
\begin{array}{cc}
\mathbf{T}^{\text{qq}} & \mathbf{1}_{\lambda} \\ 
\mathbf{1}^{\dagger}_{\lambda} & \mathbf{0}  \\
\end{array}
\right)
\left({\begin{array}{c} 
\mathbf{q}\\
\bm{\lambda} \\ 
\end{array}}\right)
=
\left(\begin{array}{c} -\bm{\chi} \\ 
\mathbf{Q}_{\alpha} 
\end{array}\right)  
\end{equation}
where $\mathbf{1}_{\lambda}$ are rectangular blocks collecting Lagrangian multipliers.

\subsection{The MCSCF/FQ approach}\label{sec:coupling}

MCSCF and FQ force field are coupled in a QM/MM fashion. The starting point is the definition of the energy of the total system:
\begin{equation}\label{eq:total-energy}
    E_{tot} = E_{\text{MCSCF}} + E_{\text{FQ}} + E^\mathrm{int}_{\text{MCSCF/FQ}}
\end{equation}
where $E_{\text{MCSCF}}$ is defined in \cref{eq:energy_cas}, and $E_{\text{FQ}}$ in \cref{eq:funct}. 

The interaction energy $E^\mathrm{int}_{\text{MCSCF/FQ}}$ is expressed in terms of the electrostatic interaction between the FQ charges and the QM solute. It can be written as: \cite{cappelli2016integrated,giovannini2020molecular}
\begin{equation}\label{eq:casscf-fq}
    E^\mathrm{int}_{\text{MCSCF/FQ}} = \sum_{i\alpha} q_{i\alpha} V_{i\alpha}(\mathbf{D})
\end{equation}
where $\mathbf{D}$ is the QM one-particle density matrix, and 
$V_{i\alpha}(\mathbf{D})$ is the QM potential acting on the charge $q_{i\alpha}$, i.e.:
\begin{equation}\label{eq:molecular-potential}
V_{i\alpha}(\mathbf{D}) = \sum_{N}^\mathrm{nuclei} \frac{Z_N}{\left|\mathbf{r}_{i\alpha}-\mathbf{R}_N\right|} - \sum_{pq}D_{pq}V^\mathrm{FQ}_{pq,i\alpha}
\end{equation}
In \cref{eq:molecular-potential}, the first term is the nuclear potential, generated by the nucleus $N$ with charge $Z_N$ located at position $\mathbf{R}_N$, whereas the second term is the electronic potential, defined as:
\begin{equation}
V^\mathrm{FQ}_{pq,i\alpha} = \Braket{\phi_p|\frac{1}{\left|\mathbf{r}-\mathbf{r}_{i\alpha}\right|}|\phi_q}    
\end{equation}
where $\mathbf{r}_{i\alpha}$ is the position of the $i$-th FQ belonging to molecule $\alpha$.

Within a SS approach, the MCSCF wavefunction is optimized with reference to a specific state $\ell$, which interacts with the polarizable environment. Thus, the total MCSCF/FQ energy functional defined in \cref{eq:total-energy} becomes: 
\begin{equation}\label{eq:energy_cas_fq}
\begin{aligned}
    E^\ell(\mathbf{D}^\ell, \mathbf{P}^\ell,  \mathbf{q}^\ell, \bm{\lambda}) = &  E_\mathrm{MCSCF}(\mathbf{D}^\ell, \mathbf{P}^\ell) \\
    & + \sum_{i\alpha} q^\ell_{i\alpha} \chi_{i\alpha} + \frac{1}{2} \sum_{i\alpha, j\beta} q^\ell_{i\alpha} T_{i\alpha, j\beta}^\mathrm{qq} q^\ell_{j\beta } + \sum_\alpha \lambda_\alpha \left[\sum_i q^\ell_{i\alpha } - Q_{\alpha}\right] \\
    &+ \sum_{i\alpha} q^\ell_{i\alpha} V_{i\alpha}(\mathbf{D}^{\ell})
\end{aligned}
\end{equation}
where $\mathbf{D}^\ell$ and $\mathbf{P}^\ell$ represent the one-particle and two-particle density matrices of the state $\ell$. FQ charges are obtained by minimizing the total functional in \cref{eq:energy_cas_fq} with respect to FQ charges and Lagrangian multipliers. Therefore, the linear system in \cref{eq:fq-linear} is modified by accounting for the QM electronic potential as an additional polarization source, i.e.:
\begin{equation}
\left(
\begin{array}{cc}
\mathbf{T}^{\text{qq}} & \mathbf{1}_{\lambda} \\ 
\mathbf{1}^{\dagger}_{\lambda} & \mathbf{0}  \\
\end{array}
\right)
\left({\begin{array}{c} 
\mathbf{q}^\ell\\
\bm{\lambda}\\ 
\end{array}}\right)
=
\left(\begin{array}{c} -\bm{\chi} \\ 
\mathbf{Q}_{\alpha} \\
\end{array}\right) + \left(\begin{array}{c} -\mathbf{V} (\mathbf{D}^\ell)\\ 
\mathbf{0} \\
\end{array}\right)
\label{eq:lin_sys}
\end{equation}
%
%
To account for mutual solute-solvent polarization, the CASSCF Hamiltonian in \cref{eq:hamiltonian-sq} is also perturbed by the presence of the FQ charges as follows: 
\begin{equation}\label{eq:ham_eff}
  \hat{H} =  \sum_{pq} \left[h_{pq} + [\mathbf{q}^\ell]^\dagger \mathbf{V}^\mathrm{FQ}_{pq}\right]  \hat{E}_{pq} + \frac{1}{2} \sum_{pqrs} g_{pqrs} \left( \hat{E}_{pq} \hat{E}_{rs} - \delta_{rq} \hat{E}_{ps} \right)
\end{equation}
where $\mathbf{V}^\mathrm{FQ}_{pq}$ is the one-electron operator defined in \cref{eq:molecular-potential}.
%
%
%
%
In addition, gradients with respect to CI and MOs coefficients defined in \cref{eq:ci_grad,eq:orb_grad}, respectively, are modified to account for the additional FQ contributions, i.e.:
%
%
\begin{align}
    g_{\mathrm{tot},K}^{(c)} & = g_{K}^{(c)} + 2 \braket{\Psi_0|\sum_{pq}[\mathbf{q}^\ell]^\dagger \mathbf{V}^\mathrm{FQ}_{pq}\hat{E}_{pq}|\Psi_K}\label{eq:ci_grad_fq} \\
    g_{\mathrm{tot},rs}^{(o)} & = 
    g_{rs}^{(o)} + 2 \braket{\Psi_0| \sum_{pq}[\mathbf{q}^\ell]^\dagger\mathbf{V}^\mathrm{FQ}_{pq} \hat{E}_{pq} |rs}\label{eq:orb_grad_fq}
\end{align}
These gradient expressions are used during the optimization of the MCSCF wavefunction. In particular, $g_{\mathrm{tot},K}^{(c)}$ is used to obtain the CI coefficients from the secular equation (see \cref{eq:ci_grad}), whereas $g_{\mathrm{tot},pq}^{(o)}$ enters the SCI eigenvalue equation for the determination of the MO coefficients (see \cref{eq:SCI}).

To summarize, an SS MCSCF/FQ calculation implies:
\begin{enumerate}
    \item computing initial $\mathbf{T}^{(0)}$ and $\mathbf{S}^{(0)}$ values;
    \item calculating the starting density matrices $\mathbf{D}^{\ell,(0)}$ and $\mathbf{P}^{\ell,(0)}$;
    \item computing the starting FQ charges $\mathbf{q}^{\ell,(0)}$ from \cref{eq:lin_sys};
    \item for $k=1,2,\dots$ until convergence:
    \begin{enumerate}
        \item $\mathbf{S}^{(k)}$ is computed from \cref{eq:ci_grad} with the inclusion of FQ contributions in \cref{eq:ci_grad_fq};
        \item $\mathbf{T}^{(k)}$ are calculated from \cref{eq:SCI} with the inclusion of FQ contributions in \cref{eq:orb_grad_fq};
        \item FQ charges  $\mathbf{q}^{\ell,(k)}$ are updated from \cref{eq:lin_sys};
        \item the SS MCSCF/FQ energy is finally computed by exploiting \cref{eq:energy_cas_fq}.
    \end{enumerate}
\end{enumerate}
The computational procedure may be generalized to SA MCSCF schemes, by taking inspiration from what has been proposed for alternative polarizable embedding methods.\cite{song2023state,song2024polarizable} 
Because such a generalization is not straightforward, this work limits the treatment to SS CASSCF wavefunctions. Extension to alternative MCSCF schemes will be presented in a future communication. 

\section{Computational Details}\label{sec:compdet}

The quality of the approach is investigated by computing vertical excitation energies of formaldehyde (FORM) and para-nitroaniline (PNA) solvated in aqueous solution. A multi-step protocol, adapted from a protocol that was specifically designed for modeling spectral signals of molecules in solution \cite{giovannini2020molecular} at the QM/MM level is exploited. 
The protocol consists of the following steps: \cite{giovannini2020molecular, gomez2022multiple}
\begin{enumerate}
    \item \textit{Definition of the system}: The QM and MM regions are defined according to the physicochemical features of the system.\cite{giovannini2020molecular} In this case, the solutes (FORM and PNA) are treated at the CASSCF level, while the polarizable FQ force field is used to represent the aqueous solvent. 
    \item \textit{Conformational sampling}: The solute-solvent phase-space is explored by using classical molecular dynamics (MD) simulations to obtain a representative conformational sampling of the solvated system. 
    Specifically, a 10 ns MD simulation of aqueous FORM (NVT) using the GROMACS package \cite{abraham2015gromacs} is performed [see section S1 in the Supporting Information (\sm) for further details].    
    In the case of PNA, two NVT MD simulations are considered: i) the PNA geometry is kept fixed in its minimum energy, as optimized at the CAM-B3LYP/aug-cc-pVDZ/PCM level of theory (MD\textsubscript{rigid}) or ii) PNA can move freely (MD\textsubscript{free}). The two MDs are taken from Refs. \citenum{ambrosetti2021quantum} (2.5 ns production run) and \citenum{giovannini2019electronic} (5 ns production run), respectively.
    \item \textit{Extraction of structures}: A set of uncorrelated snapshots is extracted from the production phase of each MD run. In particular, a single snapshot of FORM is randomly selected from the MD trajectory, while 201 snapshots of PNA are extracted from MD\textsubscript{rigid} and MD\textsubscript{free}. For each snapshot, a spherical droplet is cut. The droplet is centered on the solute with a radius large enough to account for relevant solute-solvent interactions. Specifically, the radius is 15 \AA{} for FORM and 18 \AA{} for PNA. Representative snapshots of FORM and PNA are depicted in \cref{f:structures}.
    \begin{figure}[!htbp]
        \centering
        \includegraphics[width=0.95\textwidth]{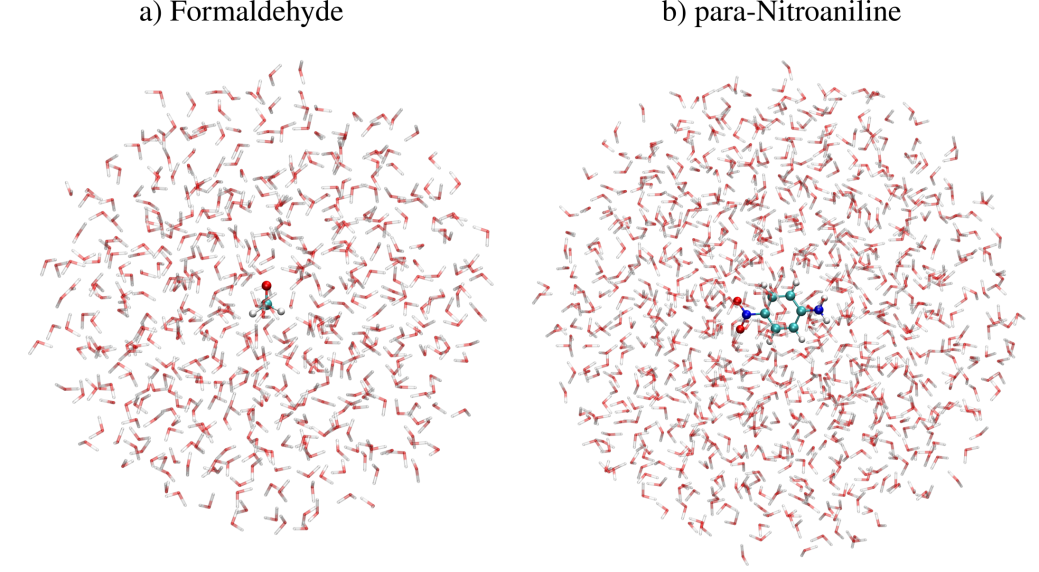} 
        \caption{Representative snapshots of formaldehyde (a) and para-nitroaniline (b) in aqueous solution.}
        \label{f:structures}
    \end{figure}
    \item \textit{QM/MM calculations}: For each spherically-shaped snapshot, vertical excitation energies are calculated at the CASSCF/FQ level using two parametrizations (FQ$^a$, taken from Ref. \citenum{rick1994dynamical} and FQ$^b$, taken from Ref. \citenum{giovannini2019effective}). To highlight the effect of solute-solvent polarization, additional calculations employing the non-polarizable ElectroStatic Potential Fitted (ESPF) \cite{ferre2002approximate} approach are performed, using TIP3P charges.\cite{jorgensen1981quantum}.

     CASSCF computed values depend on several parameters, such as the choice of starting orbitals, the active space, and the basis set.\cite{bao2019automatic,toth2016finding,roos2004relativistic,roos2005new,aquilante2020modern,li2023openmolcas} In this paper, the following procedure is exploited:
    
    \begin{itemize}
        \item Calculation of starting orbitals (at the HF\cite{roos1992multiconfigurational}  or CASSCF level) using a small basis set (ANO-RCC-MB)\cite{veryazov2011select}. This procedure permits to easily determine the MOs defining in active space; 
        \item Selection of the MOs and definition of the active space;
        \item Projection of the ANO-RCC-MB basis set into the selected, larger, basis set.\cite{aquilante2020modern, li2023openmolcas};
        \item CASSCF-SS(/classical) calculations on the ground state (GS) and excited state (ES). The vertical excitation energy is computed as the difference between CASSCF-ES and CASSCF-GS energies. In the case of QM/MM calculations, we ensure that the active space and electronic transitions are consistent for all the extracted snapshots.
    \end{itemize}

   For analyzing solvent effects and specific solute-solvent interactions, vertical excitation energies are also computed in the gas phase and by treating the solvent at the implicit PCM level\cite{tomasi2005quantum}. In both cases, FORM and PNA geometries are optimized with Gaussian16\cite{g16}, by using the B2PLYP double hybrid functional combined with the maug-cc-pVTZ-d(H) basis set.\cite{puzzarini2018diving,giovannini2019effective}  
   CASSCF/PCM calculations are performed within the non-equilibrium regime.\cite{cammi2003multiconfigurational}
   All CASSCF(/classical) calculations are performed by exploiting a locally modified version of OpenMolcas, \cite{aquilante2020modern,li2023openmolcas} where CASSCF/FQ has been implemented.
   \item \textit{Extraction of final spectra}: 
   For each system, the CASSCF/MM spectrum is obtained by averaging absorption spectra of all snapshots, after convolution with a Gaussian function with a full width at half maximum (FWHM) of 0.3 eV. 
\end{enumerate}


\section{Results and Discussion}

This section analyzes the quality of the approach to compute vertical excitation energies of FORM in aqueous solution; differences arising from the selection of the basis set, starting orbitals, active spaces, and solvent models are highlighted. Then, vertical excitation energies of PNA in aqueous solution are discussed, by highlighting the role of solvent dynamical fluctuations and comparing computed and experimental data, taken from the recent literature.\cite{kovalenko2000femtosecond}

\subsection{Formaldehyde in Aqueous Solution}\label{sec:form}

To evaluate the quality of CASSCF/FQ, the $n - \pi^*$ excitation energy of FORM in aqueous solution is selected. A single snapshot, randomly extracted from the MD trajectory, is considered. As described in Section 2, vertical excitation energies are computed as the difference between ES and GS SS-CASSCF calculations. 
Starting orbitals are determined at the HF level, at the CASSCF-SS level by optimizing the GS (CASSCF-GS), and at the CASSCF-SA level by selecting the first two states [CASSCF-SA(2]). Both a reduced (6,5) or the full valence (12,10) active spaces are tested, in line with previous studies. \cite{angeli2005casscf,angeli2005ab,hoyer2016multiconfiguration,merchan1995theoretical} Four basis sets (6-31G*, cc-pVDZ, aug-cc-pVDZ, and aug-cc-pVTZ) are selected, which allow evaluating the effect of polarization and diffuse functions on calculated excitation energies. Finally, various descriptions of the solvent environment (implicit PCM, non-polarizable ESPF, FQ with two different parametrizations - FQ$^a$ and FQ$^b$) are compared. Gas-phase values obtained at the same level (basis set, active space, and starting orbitals) are taken as a reference.

\begin{figure}[ht!]
 \includegraphics[scale=1]{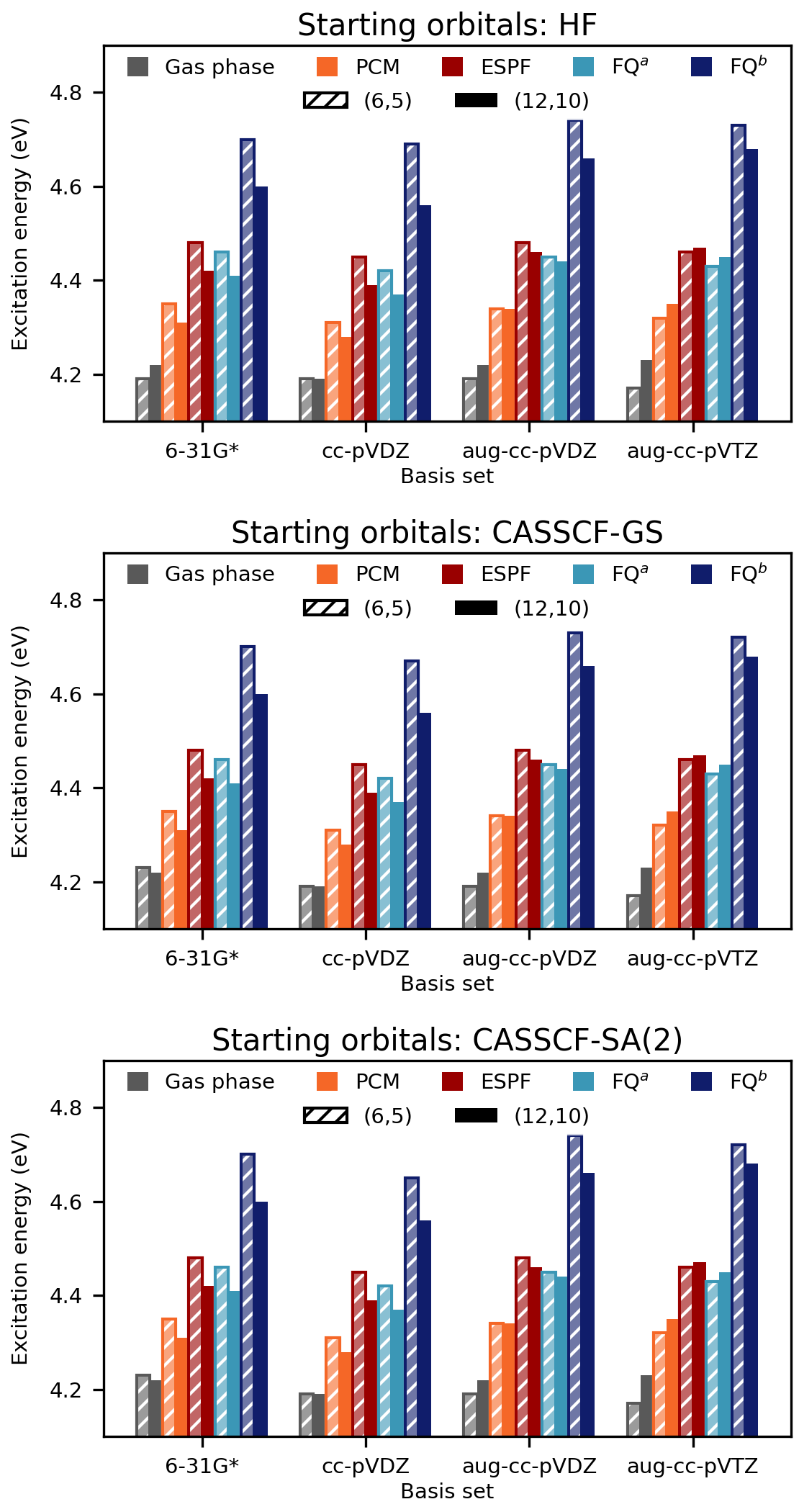} 
    \centering
    \caption{Computed FORM $n - \pi^*$ vertical excitation energies employing  HF (top panel), CASSCF-GS (middle panel), and CASSCF-SA(2) (bottom panel) starting orbitals. Selected basis set, active space, and solvation models (gas phase, PCM, ESPF, FQ$^a$, and FQ$^b$) are compared. The hatched and filled bars correspond to the (6,5) and (12,10) active spaces, respectively.}
    \label{f:benchmark_scf}
\end{figure}

\Cref{f:benchmark_scf} graphically shows computed values (raw data are reported in tables S1 to S3 in the \sm). By first focusing the attention on the effect of the selection of specific starting orbitals, 
it comes out that differences are generally negligible (0.002 eV on average). The largest, yet small, variations occur for CASSCF/FQ$^b$(6,5)/cc-pVDZ calculations. In this case, vertical excitation energies of 4.69 eV (HF), 4.67 eV (CASSCF-GS), and 4.65 eV [CASSCF-SA(2)] are obtained. This suggests that the choice of initial orbitals is not critical for aqueous FORM. 

\begin{figure*}[h]
    \centering
    \includegraphics[width=0.50\textwidth]{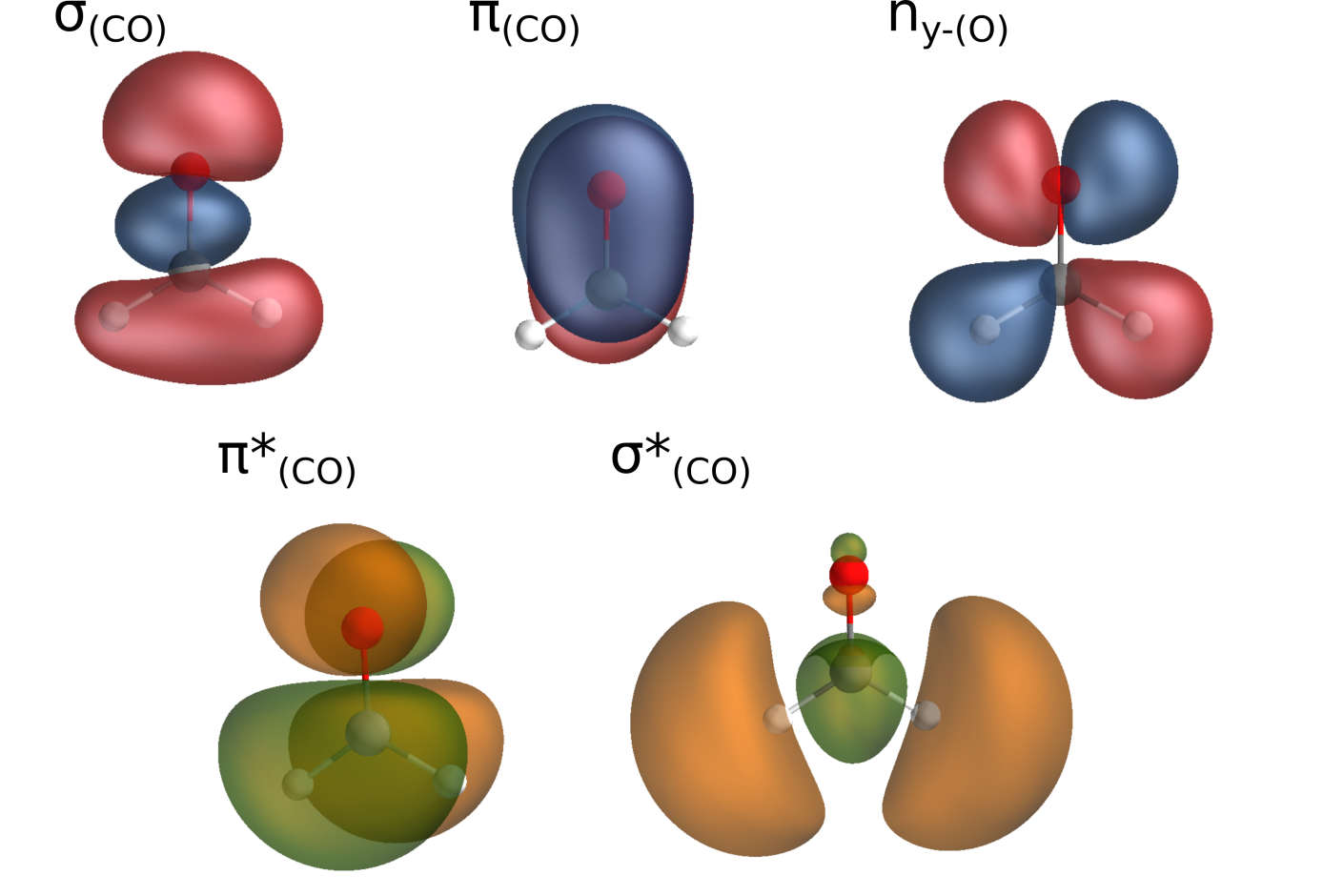}  
    \caption{HF/FQ$^b$ selected natural orbitals of aqueous FORM included in the (6,5) active space. The image illustrates $\sigma$, $\pi$, $\pi^*$, and $\sigma^*$ orbitals of the carbonyl group, and the $n_y$ orbital of oxygen. FORM lies in the $yz$ plane, with the carbonyl group aligned along the $z$ axis.}
    \label{f:orbitas_form}
\end{figure*}

We now focus on how the choice of the active space affects numerical results. The (6,5) space includes the $\sigma$, $\pi$, $\pi^*$, and $\sigma^*$ orbitals of the carbonyl group and the $n_y$ of the oxygen (see \cref{f:orbitas_form}). 
The (12,10) active space includes all MOs except for the 1s of Carbon and Oxygen atoms.
As expected, diverse excitation energies are obtained with the two active spaces, independently of the other parameters (see \cref{f:benchmark_scf}). The largest differences (0.13 eV) are reported for the FQ$^b$ parametrization combined with the 6-31G* and cc-pVDZ basis sets. 
Generally, the larger active space (12,10) yields smaller excitation energies. This is because, in the full valence active space, the ES is more stabilized than the GS, thus resulting in a lower excitation energy. Note, however, that this trend does not hold for all parameter combinations, (see \cref{f:benchmark_scf} and tables S1 to S3 in the \sm). 

Substantial differences also emerge by changing the basis set. In fact, by exploiting the (6,5) active space,  a decrease in excitation energies (0.03 eV on average) is observed by moving from the 6-31G* to cc-pVDZ basis set; values further increase moving to the aug-cc-pVDZ basis (0.03 eV on average) and finally decrease when the largest aug-cc-pVTZ basis set is employed (0.02 eV on average, see tables S1 to S3 in the \sm). When the active space is extended to (12,10), the above trends are maintained with the only exception of the aug-cc-pVTZ basis set, which yields higher excitation energies than aug-cc-pVDZ (0.01 eV on average). In general, increasing the size of the basis set stabilizes both the GS and the ES. However, depending on the basis set, such a stabilization can be larger for the GS or the ES, thus resulting in higher or lower excitation energy values, respectively. Note that the above comments yield for both gas-phase and solution and all tested solvation models. 


To analyze how computed values are affected by the selection of the solvent model, the $n - \pi^*$ transition is considered and gas phase results are taken as a reference. The inclusion of solvent effects blueshift excitation energies as compared to gas phase values. By focusing on solvatochromic shifts, i.e. the difference between the absorption energy in solution ($E^{solv}$) and gas-phase ($E^{vac}$) (see tables S1 to S3 in the \sm) it comes out that QM/PCM gives the smallest values. This is probably due to the lack of accounting for hydrogen bonding.\cite{giovannini2020molecular,giovannini2023continuum} 
In fact, all atomistic solvent descriptions yield more pronounced solvatochromic shifts. Excitation energies computed at the QM/ESPF and QM/FQ$^a$ levels are very similar and result in solvatochromic shifts that are almost twice those calculated at the QM/PCM level. 
Comparable values given by the non-polarizable ESPF and polarizable FQ$^a$ probably arise from the fact that both methods have been parametrized to reproduce the properties of bulk water. \cite{jorgensen1981quantum,rick1994dynamical} 
Notably, the largest solvatochromic shifts are obtained using the FQ$^b$ parameterization (QM/FQ$^b$), designed to reproduce solute-solvent electrostatic and polarization interactions in aqueous solutions.\cite{giovannini2019effective} Note that the largest QM/FQ$^b$ shift (0.56 eV) is obtained with the reduced active space (6,5) in combination with the aug-cc-pVTZ basis and HF starting orbitals (see table S1 in the \sm). Generally, among the various parameters considered in the above analysis, the choice of the solvent model and its parametrization are those more largely impacting computed excitation energies.

\subsection{Para-Nitroaniline in Aqueous Solution}

In this section, CASSCF/FQ is applied to the simulation of the bright $\pi-\pi^*$ excitation of PNA in aqueous solution. The solute-solvent phase space is sampled by resorting to classical MDs, from which a set of uncorrelated snapshots is extracted (see \cref{sec:compdet}). First,  the solute is kept frozen to the optimized GS geometry (MD\textsubscript{rigid}, \cref{sec:rigid}). Then, PNA is allowed to freely move (MD\textsubscript{free}, \cref{sec:flexible}). In this way, effects arising from the solvent dynamics around the solute are isolated from those following the reorganization of both PNA geometry and the solvent around it.

200 
spherically-shaped snapshots are extracted from MD\textsubscript{rigid} or MD\textsubscript{free}, then absorption energies are computed on each snapshot at the non-polarizable CASSCF/ESPF and polarizable CASSCF/FQ$^a$ and CASSCF/FQ$^b$ levels. All CASSCF calculations are performed by employing the aug-cc-pVDZ basis set and HF starting orbitals, in line with the analysis reported above for FORM (see \cref{sec:form}). 
In the case of CASSCF(12,10)/PCM calculation, the starting orbitals are obtained from CASSCF-GS(8,7)/PCM, to ensure convergence to the same electronic states for both active spaces.

\begin{figure}[h]
    \centering 
    \includegraphics[width=0.95\textwidth]{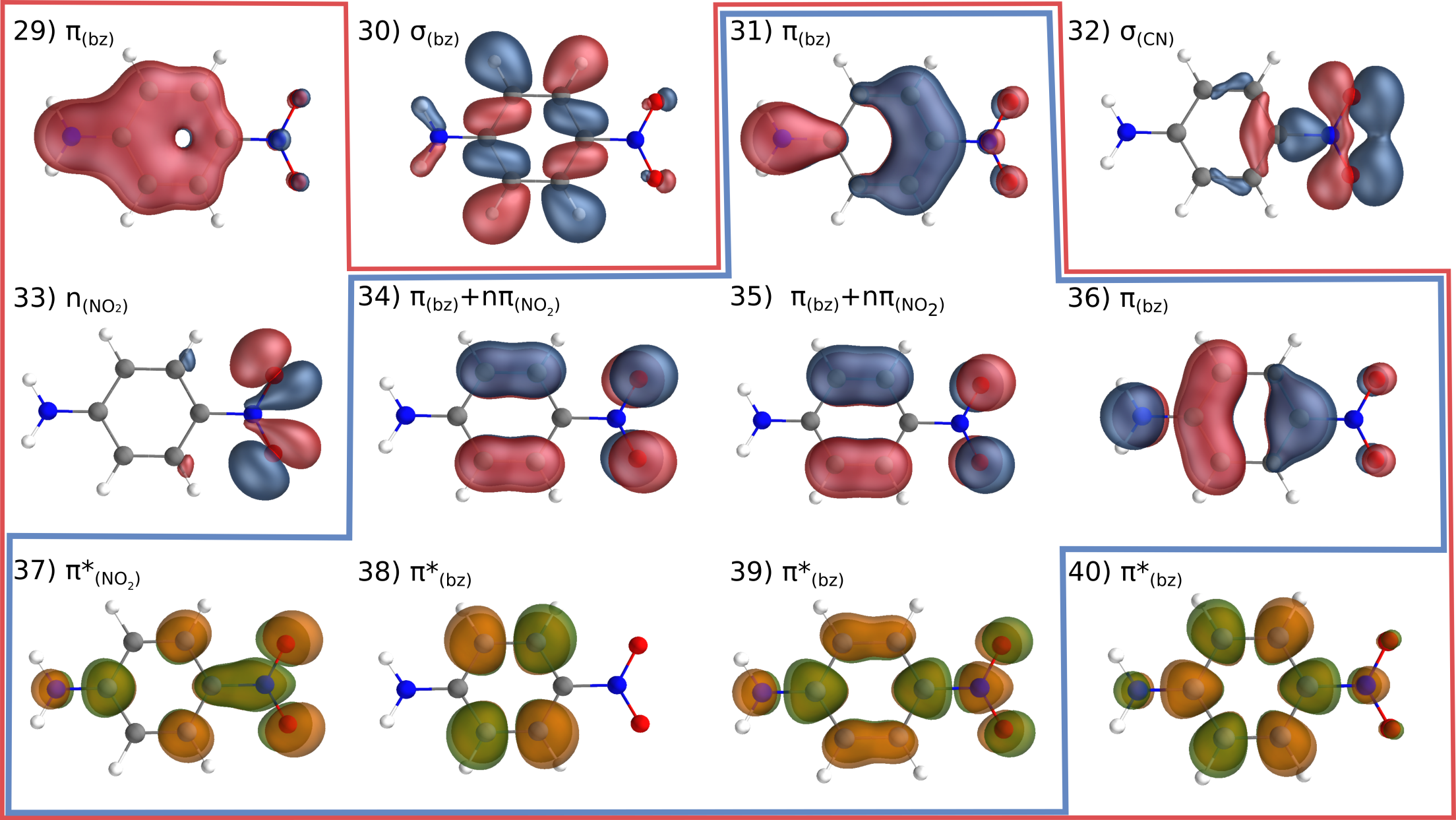}       
    \caption{HF/FQ$^b$ selected PNA natural orbitals. The image represents $\pi$ and $\pi^*$ orbitals associated with the benzene ring (bz) and the nitro group, $\sigma$ orbitals of the benzene ring and CN group and $n$ orbital of oxygen atoms of the nitro group (the labeling follows Ref. \citenum{soto2021electronic}). The blue box collects orbitals defining the (8,7) active space, while the red box defines the (12,10) active space.}
    \label{f:orbitals_pna}
\end{figure}

Two active spaces, i.e. (8,7) and (12,10), are tested for structures extracted from MD\textsubscript{rigid}, whereas only the (8,7) active space is applied to structures extracted from MD\textsubscript{free}. The (12,10) active space is selected based on Ref. \citenum{frutos2013theoretical}, while the (8,7) is chosen to investigate how the choice of the active space affects the final results. 
MOs belonging to the two active spaces and their classification are graphically reported in \cref{f:orbitals_pna}; the $\pi - \pi^*$ transition involves orbitals 36 and  37. The (12,10) active space consists of five occupied $\pi$ orbitals (29, 31, 34, 35, 36), one occupied $n$ orbital (33), and four lower-energy virtual $\pi^*$ orbitals (37, 38, 39, 40). In contrast, the (8,7) active space includes four occupied $\pi$ orbitals (31, 34, 35, 36) and three lower-energy virtual $\pi^*$ orbitals (37, 38, 39). The $\pi^*$-type orbital 40 is excluded due to its low occupation number, as indicated by the natural orbital analysis (see table S4 in the \sm).





\subsubsection{PNA/water structures extracted from MD\textsubscript{rigid}}\label{sec:rigid}


\begin{figure}[!htbp]
    \centering
    \includegraphics[scale=1]{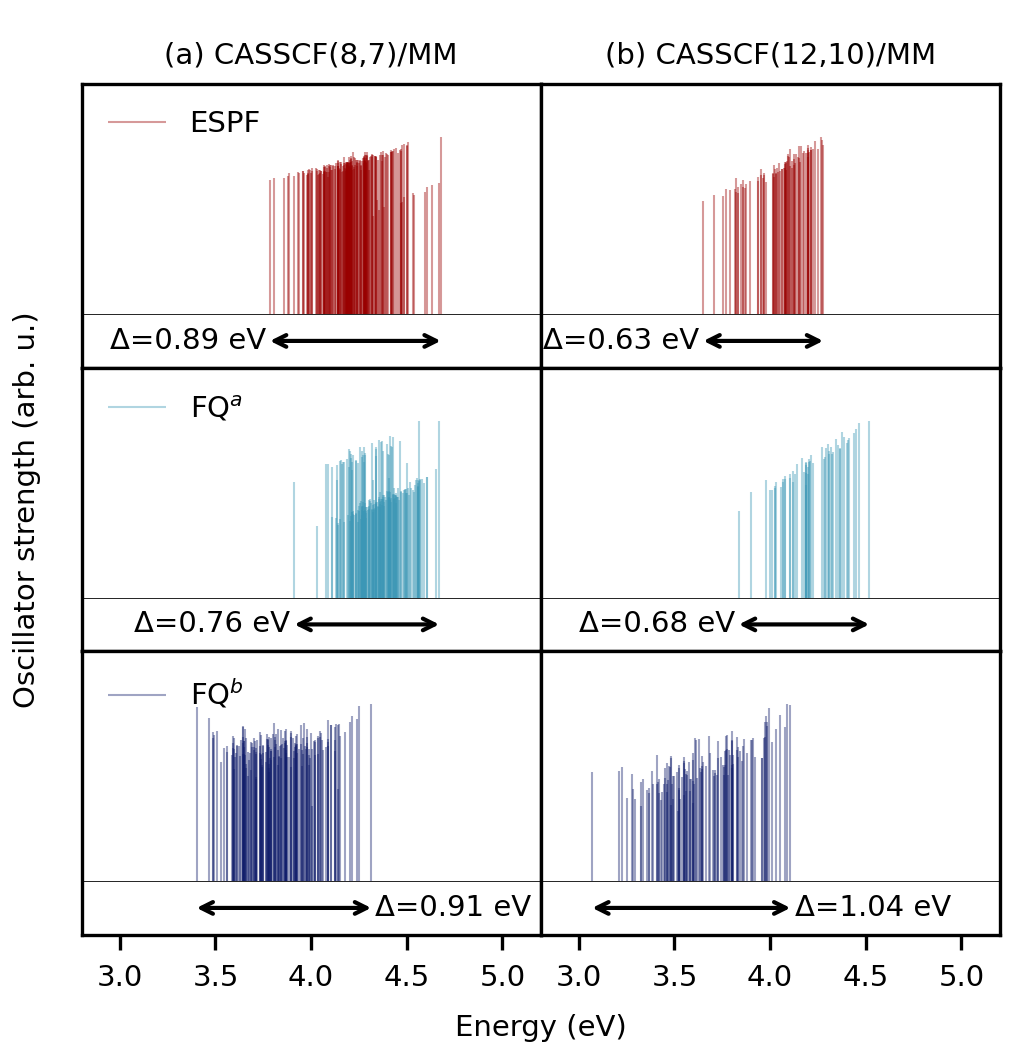}      
    \caption{Computed transitions (stick spectra) of PNA in aqueous solution (structure are taken from MD\textsubscript{rigid}) employing CASSCF/ESPF, CASSCF/FQ$^a$, and CASSCF/FQ$^b$ with active spaces (a) (8,7) and (b) (12,10). The spread in the excitation energy among all snapshots ($\Delta$) for each level of theory is also reported.}
    \label{f:stick}
\end{figure}


In \cref{f:stick}, computed CASSCF/ESPF, CASSCF/FQ$^a$, and CASSCF/FQ$^b$ excitation energies extracted from the 200 snapshots are reported as "stick spectra"; both active spaces are considered. Each vertical stick represents the excitation energy of a given snapshot with its associated oscillator strength. Clearly, both absorption energies and oscillator strengths significantly vary as a function of the snapshots, i.e. in this case from the dynamic fluctuations of water molecules around the frozen PNA molecule. Remarkably, this way of proceeding results in a solvent-induced inhomogeneous broadening of spectral bands to be naturally taken into account; since it results from solvent fluctuations, the spreading of excitation energies depends on the solvation model, and it is less accentuated for ESPF and FQ$^a$ as compared with FQ$^b$. In particular, the lowest and largest variations of absorption energies are reported for CASSCF(12,10)/ESPF (0.63 eV) and CASSCF(12,10)/FQ$^b$ (1.04 eV), thus highlighting a diverse description of solute-solvent interactions provided by the various atomistic approaches.

Before moving to comment on spectra, it is critical to ensure that the absorption signal is at convergence with respect to the number of snapshots extracted from MD simulations and exploited to compute final averaged spectra. Averaged spectra which are obtained by using an increasing number of snapshots are shown in fig. S2 in the \sm. As a measure of the statistical significance of the averaged excitation energy, table S5 in the \sm reports average excitation energies and the 95\% confidence interval. The largest error (at 95\% confidence) is reported for CASSCF(12,10)/FQ$^b$ (0.016 eV) and CASSCF(12,10)/FQ$^a$ requires the smallest number of snapshots (57) to reach convergence (error: 0.01 eV). 

\begin{figure}[!ht]
    \centering
    \includegraphics[scale=0.95]{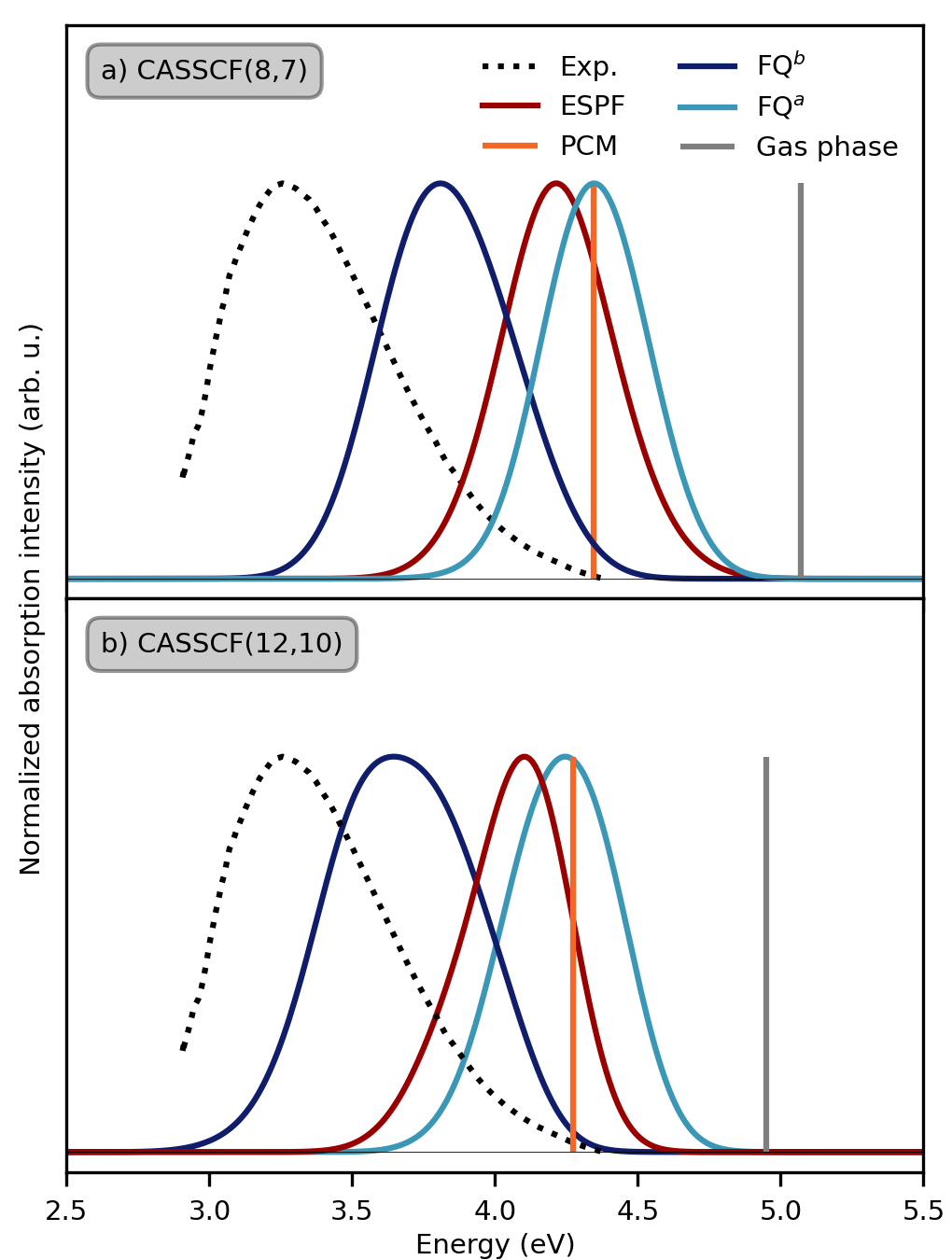}
    \caption{PNA normalized averaged absorption spectra obtained from snapshots extracted from MD\textsubscript{rigid}. (a) (8,7) and (b) (12,10) active spaces. Computed excitation energies in gas-phase and aqueous solution as obtained with the implicit PCM description are indicated. Experimental spectra taken from Ref.\cite{kovalenko2000femtosecond} are reported for the sake of comparison.}
    \label{f:spectra}
\end{figure}

\Cref{f:spectra} shows averaged absorption spectra at each CASSCF/MM level of theory for both active spaces. Absorption energies computed in the gas phase and using the implicit PCM approach at the same level of theory are also shown, along with the experimental profile reproduced from Ref. \citenum{kovalenko2000femtosecond}.

%

Moving from the gas phase (gray line) to the aqueous solution, the $\pi-\pi^\ast$ transition is red-shifted. The shift varies as a function of the solvent description and the active space. Corresponding excitation energies ($\varepsilon$) and associated solvatochromic shifts ($\delta$) for each level of theory are reported in \cref{t:shift}. 

It is also worth highlighting that the (12,10) active space gives smaller excitation energies than (8,7). This is due to a larger stabilization of the ES with respect to the GS, as also observed in the gas phase (see bottom panel of \cref{f:spectra}). As a result, the excitation energy is reduced by 0.08 eV for CASSCF/PCM, 0.11 eV for CASSCF/ESPF, 0.10 eV for CASSCF/FQ$^a$, and 0.16 eV for CASSCF/FQ$^b$ when the active space is enlarged from (8,7) to (12,10) (see \cref{t:shift}). 

We now move to compare computed results with experimental data reported in Ref. \citenum{kovalenko2000femtosecond}. From the inspection of \cref{f:spectra}, it comes out clearly that all methods overestimate the transition energy as compared to experiments, independently of the choice of the active space. This is in agreement with previous data reported in the literature and is associated with the lack of dynamic electron correlation.\cite{hoyer2016multiconfiguration} Consistently, absorption energies computed using the largest active space (12,10) are remarkably shifted towards the experimental value, with CASSCF/FQ$^b$ yielding the best reproduction of the experimental absorption band. 

To mitigate the systematic error associated with the lack of dynamic correlation, the attention can be focused on solvatochromic shifts (see \cref{t:shift}). 
The largest variation is reported for CASSCF/FQ$^b$ and CASSCF/PCM (0.04 eV), while the implicit QM/PCM yields the smallest solvatochromic shift for both active spaces. All atomistic approaches generally provide larger solvatochromic shifts, which slightly increase by passing from CASSCF/FQ$^a$ to CASSCF/ESPF. The polarizable CASSCF/FQ$^b$ method gives the largest solvatochromic shift [1.26 eV and 1.30 eV for (8,7) and (12,10) respectively], almost doubling CASSCF/FQ$^a$. CASSCF/PCM, CASSCF/ESPF, and CASSCF/FQ$^a$ underestimate the solvatochromic shift compared to the experimental value (1.0 eV),\cite{kovalenko2000femtosecond}  while CASSCF/FQ$^b$ overestimates it for both active spaces (see \cref{t:shift}). 

\begin{table}[!ht]
\begin{tabular}{l|cc|cc}
\hline
\multirow{2}{*}{Model} & \multicolumn{2}{c|}{$\varepsilon$ (eV)} & \multicolumn{2}{c}{$\delta$ (eV)}     \\
 & (8,7) & (12,10) & (8,7) & (12,10) \\
\hline
Gas phase & 5.07 & 4.95 & - & -  \\
PCM    & 4.35 & 4.27 &  0.72 & 0.68 \\
ESPF   & 4.21 & 4.10 &  0.86 & 0.85 \\
FQ$^a$ & 4.35 & 4.25 &  0.72 &  0.70 \\
FQ$^b$ & 3.81 & 3.65 &  1.26 &  1.30   \\
\hline
Exp.\cite{kovalenko2000femtosecond}  &  \multicolumn{2}{c|}{3.23}  &  \multicolumn{2}{c}{1.00} \\
\hline
\end{tabular}
\caption{Excitation energies ($\varepsilon$) and vacuo-to-water solvatochromic shifts ($\delta=\omega_\mathrm{gas} - \omega_\mathrm{solv}$) of PNA in aqueous solution (structures are extracted from MD\textsubscript{rigid}) as computed at CASSCF/aug-cc-pVDZ level of theory with the active spaces (8,7) and (12,10) and different solvent models. Experimental data taken from Ref.\citenum{kovalenko2000femtosecond} are also reported.}
\label{t:shift}
\end{table}

The diverse picture provided by the various solvent models can be explained by considering that, while ESPF and FQ$^a$ force fields are parameterized for reproducing bulk liquid water, FQ$^b$ is parameterized to reproduce electrostatic and polarization interaction energies of molecular systems in an aqueous environment. Consequently, CASSCF/FQ$^b$ predicts a larger solvent effect, resulting in an overall overestimation of the solvatochromic shift (error: 0.26/0.30 eV), while CASSCF/ESPF provides the best agreement with the experimental value (error: 0.15/0.16 eV). However, it is worth remarking that CASSCF/FQ coupling completely neglects solute-solvent non-electrostatic interactions. In particular, we have recently shown that solute-solvent Pauli repulsion is particularly relevant for consistently modeling vacuo-to-water solvatochromic shifts. \cite{giovannini2019quantum} Specifically, Pauli repulsion effects are expected to confine the QM density, overall decreasing the absolute value of solvatochromic shift,\cite{egidi2021polarizable,giovannini2019quantum} and possibly improving the agreement of CASSCF/FQ$^b$ calculation with experimental data. In parallel, the good agreement of CASSCF/ESPF with experiments might be ascribed to a favorable error cancellation (in fact, it lacks both polarization and Pauli repulsion effects).



%

%



\subsubsection{PNA/water structures extracted from MD\textsubscript{free}}\label{sec:flexible}

To achieve a more accurate description, the absorption spectrum of aqueous PNA is simulated by allowing both solute and solvent molecules to freely move (MD\textsubscript{free}). As described in Ref. \citenum{giovannini2019electronic}, during the MD PNA oscillates predominantly around its planar configuration, thus indicating that the benzene ring is substantially rigid. 
The dihedral distribution functions of the three main dihedral angles, as shown in Fig. 5 of Ref. \citenum{giovannini2019electronic}, reveal that the nitro group exhibits rotational freedom, while the NH$_2$ group is confined to oscillations around its equilibrium geometry.



%
As already mentioned, the (8,7) active space is exploited. This choice is guided by two practical aspects. First, as shown in \cref{sec:rigid}, the solvatochromic shift slightly depends on the choice of the active space, therefore we can speed up the computation by reducing the active space to the minimum required to study the $\pi-\pi^*$ transition. Second, given the conformational flexibility of PNA, MOs undergo significant mixing and changes across different conformations, making the consistent selection of the (12,10) active space challenging (see fig. S4 in the \sm for an example of mixed orbitals in a random snapshot).

\begin{figure}[!ht]
    \centering
 \includegraphics[scale=1]{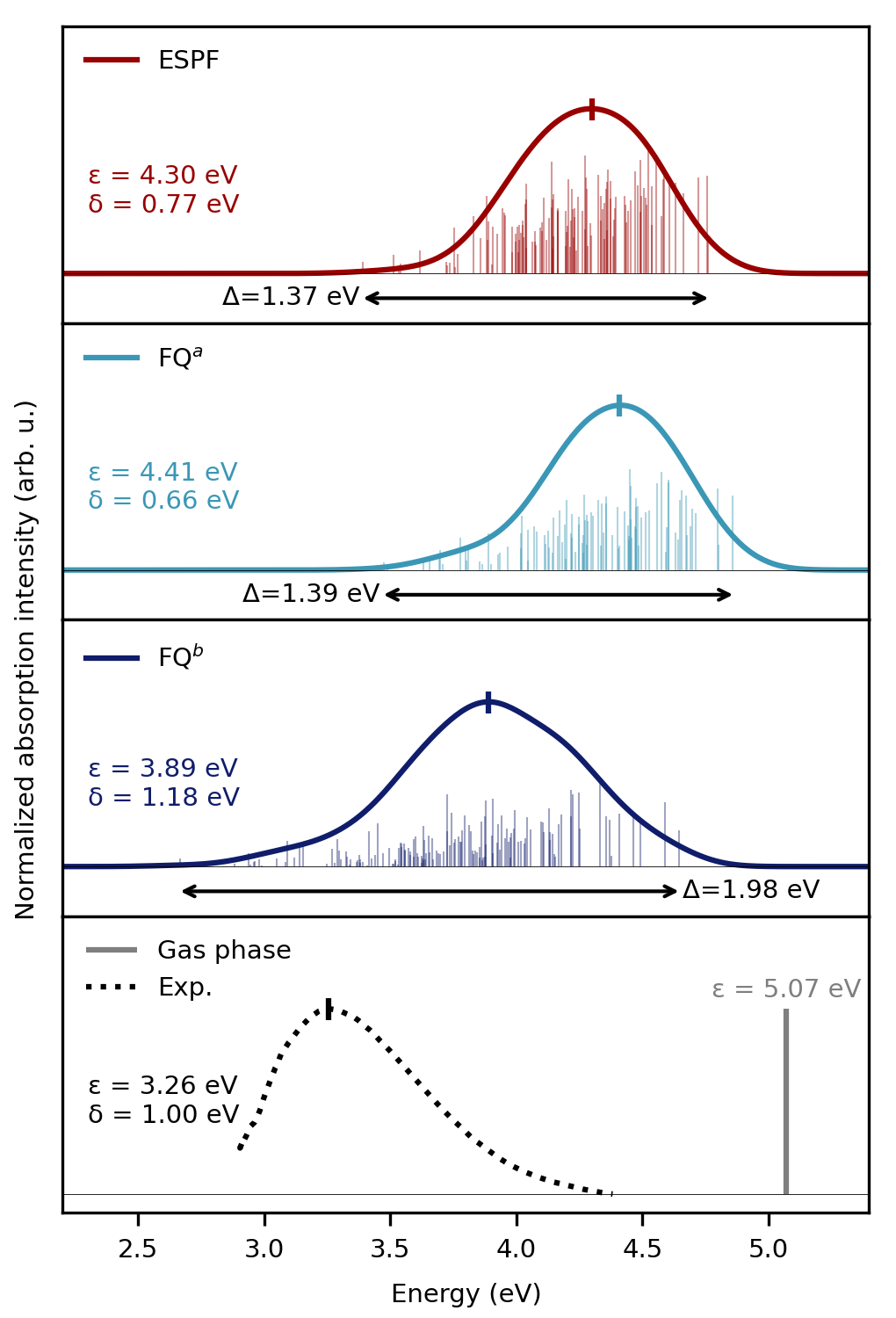}
    \caption{PNA normalized stick and absorption spectra in aqueous solution obtained with different solvent models and by using the (8,7) active space. Structures are extracted from MD\textsubscript{free}. The intensity of each stick is multiplied by a factor of 5 for ease of visualization. Maximum excitation energy ($\varepsilon$) and solvatochromic shift ($\delta$) in eV are reported in each panel. Gas-phase values and experimental spectra \cite{kovalenko2000femtosecond} are shown in the lowest panel.}
    \label{f:spectra-flex}
\end{figure} 

In \cref{f:spectra-flex}, stick and averaged spectra, average excitation energies ($\varepsilon$), and solvatochromic shifts ($\delta$) for each CASSCF(8,7)/MM level of theory are reported. Notably, PNA geometrical rearrangement significantly impacts the variability of the absorption energy among sampled conformations, almost doubling the spreading of excitation energy ($\Delta$) as compared to MD\textsubscript{rigid} (see \cref{f:stick}). 
To investigate the origin of CASSCF/MM inhomogeneous broadening observed in \cref{f:spectra-flex}, CASSCF/ESPF, CASSCF/FQ$^a$, and CASSCF/FQ$^b$ excitation energies are analyzed as functions of the three PNA main dihedral angles (see fig. S5 in the \sm). This analysis aims to determine whether the observed peak broadening is due to PNA geometrical fluctuations or solute-solvent interactions. The variability of excitation energies is quite similar for the three CASSCF/MM approaches, showing no correlation with dihedral angles. This indicates that the different broadening is primarily due to the description of solvent molecules arising from the three CASSCF/MM models. 

The convergence of spectra as increasing the number of snapshots is analyzed in fig. S3 in the \sm. Although the convergence is slower compared to MD\textsubscript{rigid}, the number of selected snapshots is sufficient to converge the final excitation energy with a maximum error of 0.077 eV in the case of CASSCF/FQ$^b$ calculations. 

As for MD\textsubscript{rigid}, final absorption profiles are obtained by convoluting each stick spectrum with a Gaussian lineshape with a FWHM of 0.3 eV; the results are reported in \cref{f:spectra-flex}. Clearly, the broadening of absorption peaks increases, as it was previously observed for the stick distributions. The relative positions of FQ$^b$, FQ$^a$, and ESPF absorption maxima are consistent with MD\textsubscript{rigid} (see \cref{f:spectra}). Furthermore, excitation energies in \cref{f:spectra-flex} exhibit only a slight blueshift with respect to \cref{t:shift},: 0.09 eV for ESPF, 0.06 eV for FQ$^a$ and 0.08 eV for FQ$^b$.

Given the small increment in excitation energies, solvatochromic shifts are reduced with respect to MD\textsubscript{rigid}, as it is shown by comparing \cref{f:spectra-flex} and \cref{t:shift}. As a consequence, solvatochromic shifts obtained using CASSCF/ESPF and CASSCF/FQ$^a$ are strongly underestimated, whereas CASSCF(8,7)/FQ$^b$ is in the best agreement with the experimental value, yielding a value of 1.18 eV. As discussed in the previous section, the inclusion of non-electrostatic interactions and dynamic correlation may bring the CASSCF/FQ$^b$ solvatochromic shift closer to the experimental value. 

\section{Summary and future perspectives}

A fully atomistic multiscale model based on the coupling between an MCSCF Hamiltonian and the polarizable FQ force field has been developed within a state-specific approach. Fluctuations in calculated values, arising from the structural rearrangement of solute and solvent, have been recovered by averaging computed values on hundreds of structures generated from a conformational sampling performed by running MD simulations (by either keeping the solute frozen or allowing it to freely move). The computational protocol has required meticulous attention to ensure the consistency of the active space across the set, and also a careful inspection of the consistency of the modeled transition. 

The approach and the associated computational protocol have been validated by computing the $n-\pi^*$ transition of aqueous formaldehyde and the $\pi-\pi^*$ transition of aqueous para-nitroaniline. As expected, computed excitation energies and vacuo-to-solvent solvatochromic shifts vary with different choices of active space. However, even larger modifications are observed as a function of the selected solvation model (continuum vs explicit, non-polarizable, or polarizable ). In the case of PNA, a good agreement between computed and experimental solvatochromic shifts is obtained by using the CASSCF(8,7)/FQ$^b$ method and allowing the solute and solvent to rearrange during the MD run employed for the conformational sampling.

The reported method has some limitations, which will be overcome in future works. First, dynamic correlation can be recovered, for instance by extending the approach to the second-order perturbation theory with a CASSCF reference wavefunction, as in the CASPT2 model. \cite{andersson1990second} 

Additionally, the state-specific approach that has been exploited here may result unphysical when dealing with multiple states simultaneously, e.g. when the model is coupled to methods to describe conical intersections and more generally non-adiabatic dynamics. In these cases, a state-average formulation appears more physically consistent.\cite{song2023state}

From the purely numerical point of view, the accuracy of the method could be further enhanced by developing on-purpose FQ parametrizations based on CASSCF reference data. In fact, FQ$^a$ and FQ$^b$ parametrization, which have been tested in the numerical segment of this study, have been developed and tested for the calculation of spectral properties at DFT level.

Moreover, similarly to any other classical electrostatic force field, FQ lacks the description of non-electrostatic interactions, which could play a significant role, especially in combination with very accurate wavefunctions. 
A possible way to account for such interactions would be to extend to MCSCF approaches the method that has been developed recently by some of the present authors,\cite{giovannini2017general} and which formulates Pauli repulsion and dispersion in terms of the QM density. As an alternative, an intermediate layer between the CASSCF and FQ regions could be exploited, thus recovering a fully quantum interaction between the CASSCF region and the first solvation shells, and resorting to FQ to describe long-range electrostatic interactions, in line of previous works of some of us.\cite{egidi2021polarizable,lafiosca2023multiscale}

\begin{acknowledgement}
We gratefully acknowledge the Center for High-Performance Computing (CHPC) at SNS for providing the computational infrastructure.
\end{acknowledgement}

\begin{suppinfo}
Computational details and analysis of MD simulations of FORM in aqueous solution. Raw excitation energies of FORM in aqueous solution by varying the starting orbitals, active space, basis set and solvent description. Occupation numbers of PNA natural orbitals. Convergence analysis for CASSCF/MM calculations on PNA snapshots extracted from MD\textsubscript{rigid} and MD\textsubscript{free}. 
Natural orbitals of a random PNA snapshot extracted from MD\textsubscript{free}. Correlation plot of excitation energies and PNA dihedral angles for CASSCF(8,7)/MM calculations.
\end{suppinfo}

\bibliography{biblio,references}

\begin{mcitethebibliography}{95}
\bibitem[Lin and Truhlar(2007)Lin, and Truhlar]{lin2007qm}
Lin,~H.; Truhlar,~D.~G. QM/MM: what have we learned, where are we, and where do
  we go from here? \emph{Theor. Chem. Acc.} \textbf{2007}, \emph{117},
  185--199\relax
\mciteBstWouldAddEndPuncttrue
\mciteSetBstMidEndSepPunct{\mcitedefaultmidpunct}
{\mcitedefaultendpunct}{\mcitedefaultseppunct}\relax
\EndOfBibitem
\end{mcitethebibliography}

\end{document}